\documentclass[twocolumn,notitlepage,prl,superscriptaddress,showpacs]{revtex4-1}
\usepackage{hyperref}
\usepackage[usenames,dvipsnames]{color}
\usepackage{amsmath}
\usepackage[english]{babel}
\usepackage{amssymb}
\usepackage{graphicx}
\usepackage{epstopdf}
\usepackage{tabularx}
\usepackage[]{todonotes}

\begin{document}

\title{
Magnetic Correlations in the Two-dimensional Repulsive Fermi Hubbard Model
}

\author{Fedor \v{S}imkovic IV.}
\affiliation{Department of Physics, King's College London, Strand, London WC2R 2LS, UK}
\author{Youjin Deng}
\affiliation{Hefei National Laboratory for Physical Sciences at Microscale and Department of Modern Physics, University of Science and Technology of China, Hefei, Anhui 230026, China}
\author{N. V. Prokof'ev}
\affiliation{Department of Physics, University of Massachusetts, Amherst, MA 01003, USA}
\affiliation{National Research Center ``Kurchatov Institute,'' 123182 Moscow, Russia}
\author{B. V. Svistunov}
\affiliation{Department of Physics, University of Massachusetts, Amherst, MA 01003, USA}
\affiliation{National Research Center ``Kurchatov Institute,'' 123182 Moscow, Russia}
\author{I. Tupitsyn}
\affiliation{Department of Physics, University of Massachusetts, Amherst, MA 01003, USA}
\affiliation{National Research Center ``Kurchatov Institute,'' 123182 Moscow, Russia}
\author{Evgeny Kozik}
\affiliation{Department of Physics, King's College London, Strand, London WC2R 2LS, UK}

\date{\today}

\begin{abstract}
The repulsive Fermi Hubbard model on the square lattice has a rich phase diagram near half-filling (corresponding to the particle density per lattice site $n=1$):
for $n=1$ the ground state is an antiferromagnetic insulator, at $0.6 < n \lesssim 0.8$,
it is a $d_{x^2-y^2}$-wave superfluid (at least for moderately strong interactions $U \lesssim 4t$ in terms of the hopping $t$),
and the region $1-n \ll 1$ is most likely subject to phase separation. Much of this physics is preempted at finite temperatures and to an extent driven by strong magnetic fluctuations, their quantitative characteristics and how they change with the doping level being much less understood. Experiments
on ultra-cold atoms have recently gained access to this interesting fluctuation regime, which is now under extensive investigation. In
this work we employ a self-consistent skeleton diagrammatic approach to quantify the
characteristic temperature scale $T_{M}(n)$ for the onset of magnetic fluctuations with a
large correlation length and identify their nature. Our results suggest that the strongest fluctuations---and hence highest $T_{M}$ and easiest experimental access to this regime---are observed at $U/t \approx 4-6$.
\end{abstract}


\maketitle


The fermionic Hubbard model \cite{hubbard1963electron,anderson1963theory,anderson1997theory,Lee2006},
defined by the square lattice Hamiltonian
\begin{align}
	H = -  t\sum_{\left<i,j\right> \sigma}  \hat{c}^{\dagger}_{i, \sigma} \hat{c}_{j, \sigma}
+ U \sum_{i} \hat{n}_{i, \uparrow} \hat{n}_{i \downarrow} -\mu \sum_{i, \sigma} \hat{n}_{i, \sigma}
    \label{Hubbard}
\end{align}
has for years played a crucial role in studies of correlated electrons in solids;
it is regarded as one of the ``standard models'' of condensed matter physics to introduce
and discuss Mott insulating phases, antiferromagnetic (AFM) correlations, novel mechanisms of
superconducting pairing, non-Fermi-liquid behavior, etc.
In Eq.~(\ref{Hubbard}) and in what follows, the nearest-neighbor hopping amplitude $t$
is set to be the energy and temperature unit (distances are measured in units of the
lattice constant);
$U$ is the on-site repulsive coupling constant;
$\mu$ is the chemical potential; $\hat{c}^{\dagger}_{i, \sigma}$ and $\hat{c}_{i, \sigma}$
create and annihilate (respectively) a fermion of the spin component $\sigma \in \{\uparrow, \downarrow\}$
on the site $i$, and $\hat{n}_{i, \sigma}$ counts the number of fermions of a particular spin on a given lattice site.

\begin{figure}[htbp]
\vspace*{0.0cm}
\includegraphics[scale=0.4,angle=-90,width=0.95\columnwidth]{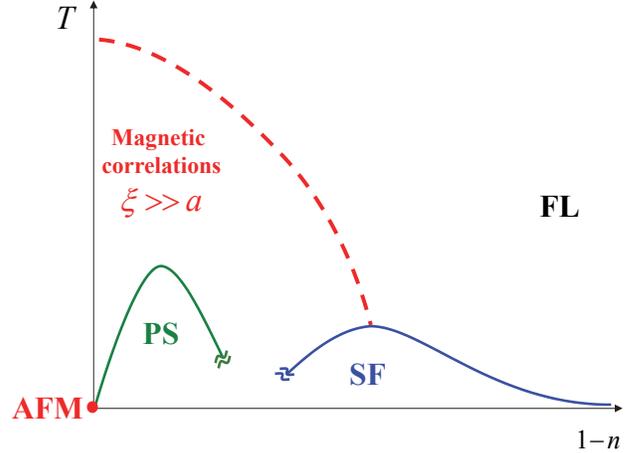}
\vspace*{0.0cm}
\caption{\label{fig1} (color online). Schematic phase diagram of the repulsive Fermi-Hubbard
model on the square lattice near half-filling, showing the Fermi liquid (FL), superfluid (SF),
and phase separation (PS) regimes, as well as the antiferromagnetic ground state (AFM) at half-filling.
Strong magnetic correlations with large correlation length $\xi \gg a$ are observed below the
dashed red line.
}
\end{figure}

On the one hand, Eq.~(\ref{Hubbard}) involves a number of crucial simplifications that make
it qualitatively different from real materials, such as high-$T_c$ superconductors.
The most important ones include
(i) strictly two dimensional (2D) as opposed to the strongly anisotropic 3D geometry,
(ii) neglect of long-range Coulomb interactions,
(iii) suppression of hopping matrix elements beyond the nearest-neighbor ones ($t'=0$),
(iv) single-band approximation, and
(v) absence of electron-phonon coupling.
Correspondingly, the model (\ref{Hubbard}) cannot feature an ordered AFM phase at a non-zero temperature,
but is allowed to have a first-order transition between phases with different electron densities,
not to mention that $t'=0$ leads to the Fermi surface nesting and particle-hole symmetry at
$n=\left<n_{i\uparrow} + n_{i\downarrow} \right>=1$.
As a result, the schematic phase diagram of (\ref{Hubbard}) in the doping-temperature plane
shown in Fig.~\ref{fig1} (see discussion below)
is distinct from the ``canonical" picture of high-$T_c$-type materials \cite{Lee2006}.
On the other hand, advances in ultra-cold atomic experiments have
made it possible to accurately emulate the model (\ref{Hubbard}) on optical lattices
\cite{jaksch1998cold, lewenstein2007ultracold, kohl2005fermionic,jordens2008mott, schneider2008metallic,hulet2015antiferromagnetism, greif2015formation,parsons2016site,greinerprivate}, bringing ultra-cold atom experiments into the region of the phase diagram (Fig.~\ref{fig1}),
where they can cooperate with the state-of-the-art numerical methods to reveal the underlying
physics. Numerical results can now be directly compared
to experiments and {\it vice versa} dramatically increasing the importance of producing
reliable data sets.

Recent years have seen a remarkable progress in unveiling the $T=0$ phase diagram of the Hubbard model (\ref{Hubbard}).
Well-understood regions include the limit of vanishing densities $n\rightarrow 0$
\cite{kagan1989increase, baranov1992d, chubukov1992pairing, chubukov1993kohn, zanchi1996superconducting, halboth2000d, fukazawa2002third} and vanishing interaction strength
\cite{hlubina1999phase,raghu2010superconductivity,romer2015pairing, simkovic2016ground}.
For densities $n \le 0.7$ and coupling strength $U \le 4$, the ground state is a BCS
superfluid (with the $d_{x^2-y^2}$-wave symmetry at density $n>0.6$) \cite{deng2015emergent}.
At half-filling $n=1$, the ground state is an AFM insulator for any $U$
\cite{hirsch1985two,hirsch1989antiferromagnetism,white1989numerical,furukawa1992two,varney2009quantum}.
Being a qualitative property, AFM order can only disappear (with doping) by a quantum phase transition, with
the simplest scenario being that of phase separation (PS). The first PS state was proposed
to be a mixture of AFM and ferromagnetic (FM) orders in the region
of small doping $\delta =1-n\ll 1$ and large $U$ \cite{penn1966stability, nagaoka1966ferromagnetism,tasaki1998nagaoka};
this conjecture has been later supported numerically for $U>25$ \cite{zitzler2002magnetism}.
The instability of the model towards incommensurate AFM and domain wall formation was also
reported in Refs.~\cite{schulz1990incommensurate,lin1991ground}. Recently, PS for small values of $U$
was observed in Auxiliary-field Quantum Monte Carlo \cite{chang2008spatially,sorella16}
and variational \cite{tocchio16} studies.

Much less is known conclusively about the finite-temperature behavior.
Given that the correlation length $\xi$ for AFM correlations at $n=1$ diverges exponentially
fast when $T\to 0$, there exists a relatively high temperature $T_{M}$ below which magnetic
correlations extend over many lattice sites and electronic degrees of freedom are getting
locked in collective modes. The characteristic temperature scale $T_{M}(n)$
is supposed to decrease with increasing the doping level. Since magnetic correlations and
fluctuations are considered to be the prime reason for PS and BCS phenomena near $n=1$,
quantifying their behavior is paramount to understanding the finite-$T$ phase diagram
of the Fermi Hubbard model in 2D.

\begin{figure}
\centering
\includegraphics[width=0.95\linewidth]{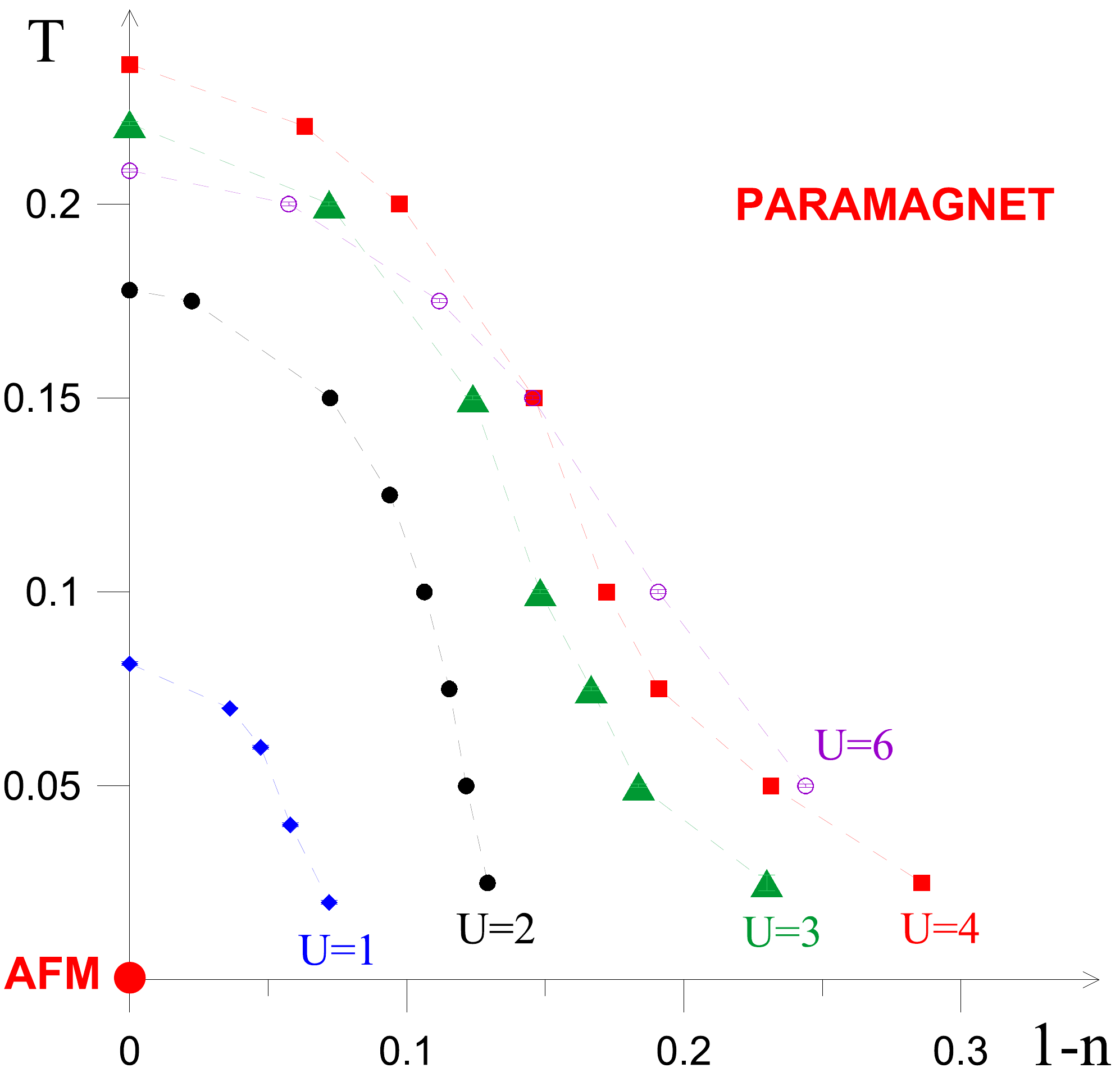}
\caption{\label{fig2} (color online). Onset of strong (incommensurate) magnetic fluctuations
as calculated by the lowest-order GGGW method (see text) for various values of $U$.
The region of strong fluctuations increases with $U$ and reaches its maximum at around $U=4$
at $n=1$.
}
\end{figure}

In this Letter, we employ a skeleton diagrammatic approach to
quantify the $T_{M}(n )$ scale, see Fig.~\ref{fig2}, and shed light on the structure of
dominant fluctuations. Experiments with ultra-cold $^6\text{Li}$ fermions have now
reached temperatures low enough to directly observe magnetic correlations in the model (\ref{Hubbard})
\cite{hulet2015antiferromagnetism, greif2015formation,parsons2016site}, with the
lowest temperatures attained on the scale of $T \sim 0.25$ \cite{greinerprivate}.
%
\begin{figure}[htbp]
\vspace*{0.0cm}
\includegraphics[scale=0.4,angle=-90,width=0.95\columnwidth]{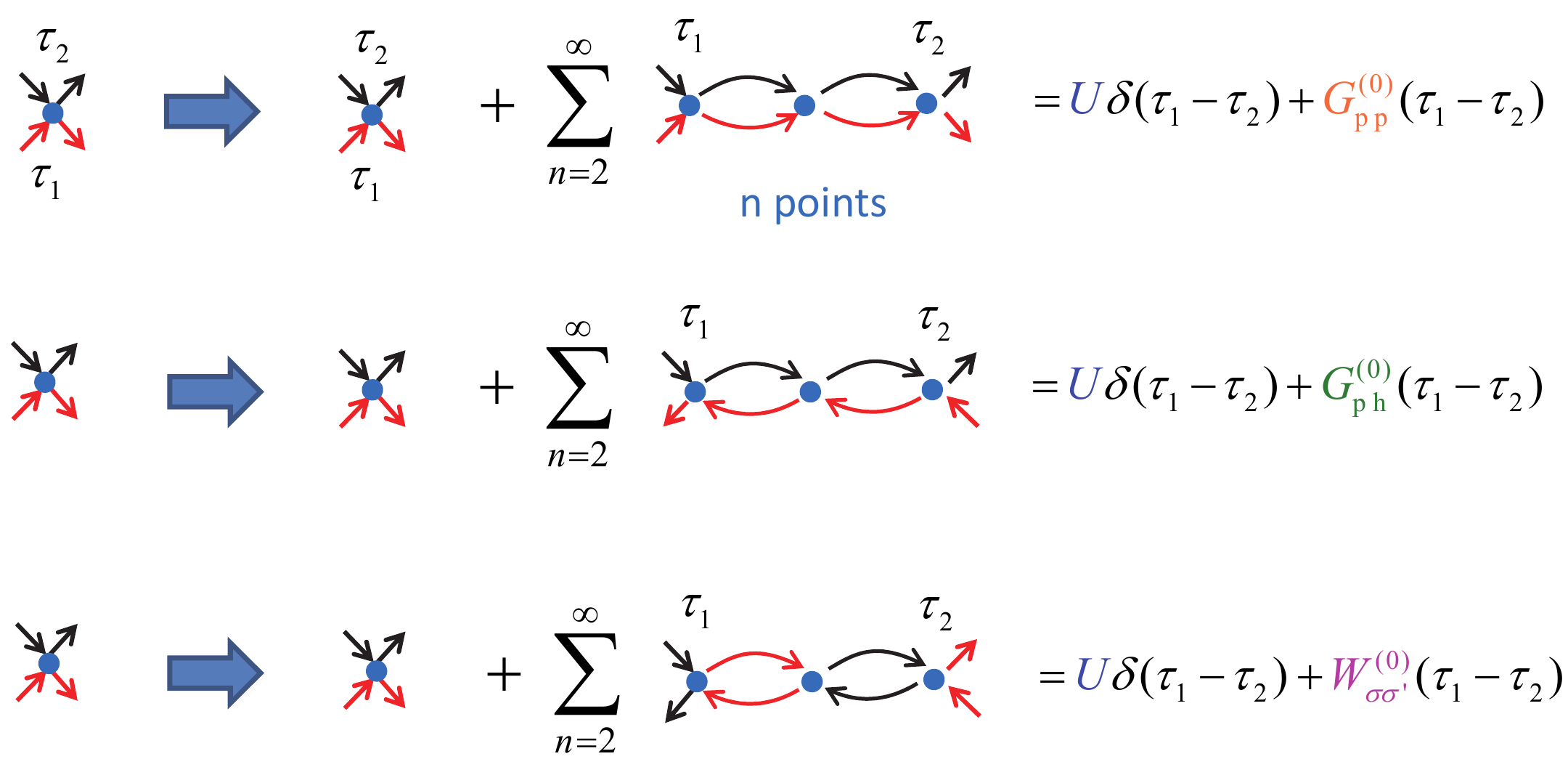}
\vspace*{0.0cm}
\caption{\label{fig3} (color online). Simple geometric series constructed by connecting
bare interaction vertexes using various pairs of bare single-particle propagators
(Green's functions) $G^{(0)}_{\sigma}$. Red and black colors stand for spin-up and
spin-down propagators.
}
\end{figure}
{\it Method.}
The imaginary-time spin correlation function $\chi (\tau,i) = \langle S_z(\tau , i ) S_z(0,0) \rangle $
and its Fourier transform $\chi (\omega_m, \vec{q})$ at bosonic
Matsubara frequencies $\omega_m =2\pi T m$ ($m$ is an integer) was computed within the skeleton diagrammatic framework based on self-consistently renormalised (``dressed'') elements in four separate channels: particle-particle and particle-hole pair propagators, screened interaction, and single-particle propagator.

\begin{figure}[htbp]
\vspace*{0.0cm}
\includegraphics[scale=0.4,angle=-90,width=0.95\columnwidth]{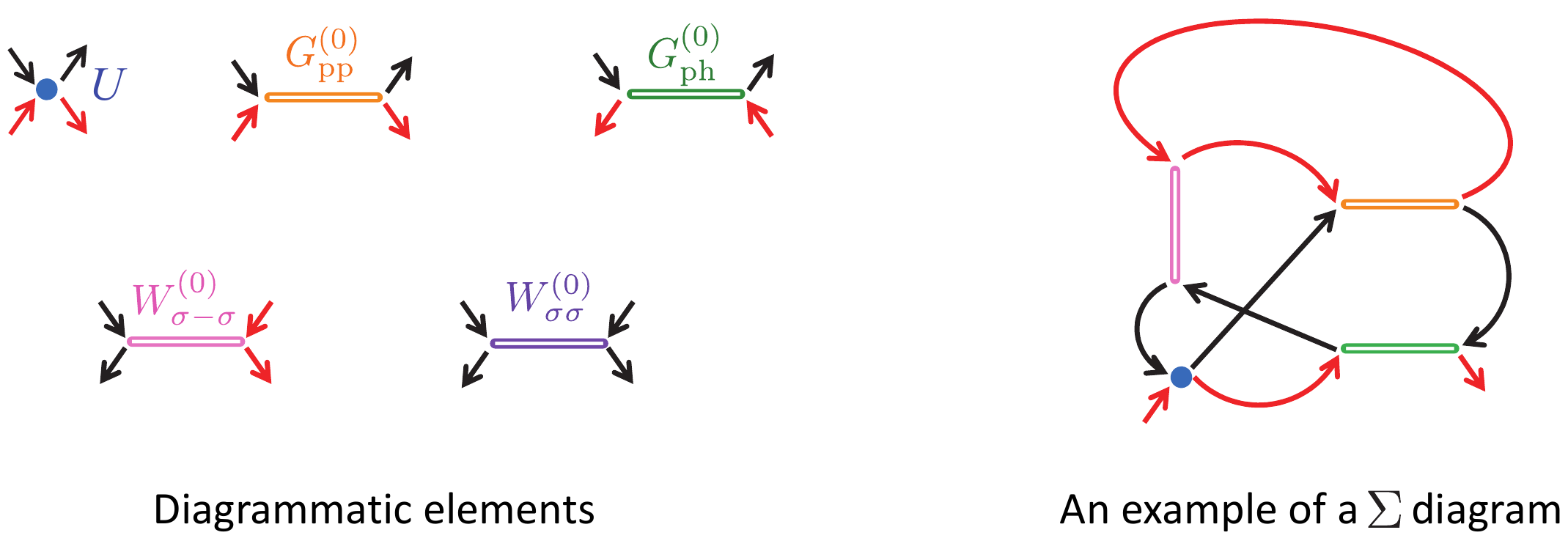}
\vspace*{0.0cm}
\caption{\label{fig4} (color online). Diagrammatic elements based on ladders and screened interactions
and an example of a particular diagram for the single-particle self-energy $\Sigma $ based on them.
 }
\end{figure}

To define the framework, we first note that in the standard weak-coupling expansion in powers of $U$ (for introduction see, e.g., \cite{AGD}) every instance of the bare interaction vertex can be replaced by either of the three types of infinite sums shown in Fig.~\ref{fig3}. These sums originate from three possible ways of connecting bare vertexes by non-interacting Green's functions $G^{(0)}_{\sigma}$ to form a geometric series and are commonly referred to as particle-particle ($G_{pp}^{(0)}$), particle-hole ($G_{ph}^{(0)}$), and bubble (screened interaction $W_{\sigma \sigma'}^{(0)}$) ladders. Note that the functions $G_{pp}^{(0)}$, $G_{ph}^{(0)}$, and $W_{\sigma \sigma'}^{(0)}$, have the same structure as the single-particle propagators $G_{\sigma}^{(0)}$; i.e., thanks to the local nature of the Hubbard interaction $U$, they depend only on one lattice coordinate and time. Therefore, they are represented diagrammatically as lines, whereas the bare vertex is a point \footnote{Bare interaction needs to be treated separately to define precise rules for avoiding double counting.}; the complete set of elements is shown schematically in Fig.~\ref{fig4} (left panel). An arbitrary diagram can now be constructed
by taking any number of these elements and connecting their incoming
and outgoing ends with propagator lines, as exemplified in the the right panel of Fig.~\ref{fig4}. The resulting series contains significantly fewer terms because a large fraction of diagram topologies in the weak-coupling expansion are accounted for by the ladder sums.
%

The standard step that allows further reduction of the diagrammatic space is replacing all the bare lines with the ones dressed by an infinite geometric series of all sorts of diagrammatic insertions, referred to as self-energies, and dropping a substantial fraction of diagram topologies that are double-counted as a result. This leads to the self-consistent so-called \textit{skeleton} technique \cite{AGD}, where one computes the diagrammatic series for the self-energies $\Sigma_{\sigma}$, $\Sigma_{pp}$,  $\Sigma_{ph}$ and $\Pi_{\sigma \sigma'}$ constructed from some approximation to the dressed lines $G_{\sigma}$, $G_{pp}$, $G_{ph}$, $W_{\sigma \sigma'}$ and obtains the next approximation by solving the corresponding Dyson equations:
\begin{equation}
G_{\sigma}=\frac{G_{\sigma }^{(0)} }{1-G_{\sigma }^{(0)} \Sigma_{\sigma }} \;, \;\;\;\;
W=\frac{U}{1-U \Pi } - U  \;,
\label{Dysons1}
\end{equation}
\begin{equation}
G_{pp} =\frac{U}{ 1-U \Sigma_{pp} } - U  \;,    \;\;\;\;
G_{ph} =\frac{U} { 1-U \Sigma_{ph} } - U \;,
\label{Dysons2}
\end{equation}
which are simple algebraic relations in the momentum/frequency representation. The diagrammatic calculation of self-energies built on the solutions of Eqs.~(\ref{Dysons1}), (\ref{Dysons2}) for all the diagram lines is repeated iteratively until convergence. The spin-spin correlation function is then directly related to the polarization $\Pi$ by
\begin{equation}
\chi (\omega_n,\vec{q}) = \operatorname{Tr} \, S_z \frac{\Pi(\omega_n,\vec{q})}{1-\Pi(\omega_n,\vec{q})U } S_z \, ,
\label{chipi}
\end{equation}
with the trace taken over the spin index. This scheme can be abbreviated as \textit{GGGW} to emphasise the four renormalization channels.

\begin{figure}[htbp]
\includegraphics[scale=0.4,angle=-90,width=0.95\columnwidth]{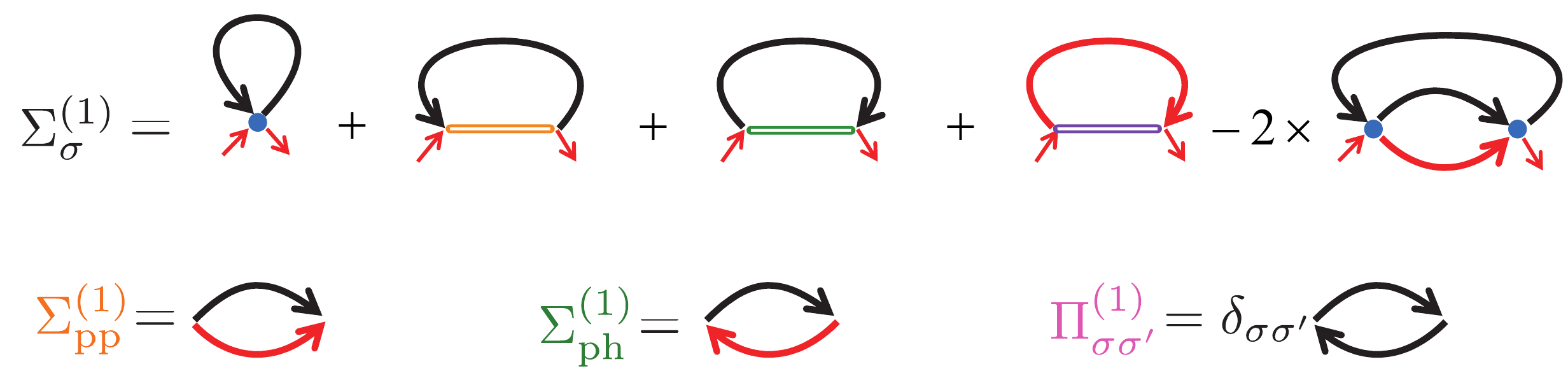}
\vspace*{0.0cm}
\caption{\label{fig5} (color online). First-order skeleton graphs for all the self-energies evaluated in this work. The dressed lines in the diagrams are determined self-consistently by Eqs.~(\ref{Dysons1}), (\ref{Dysons2}).
 }
\end{figure}

To avoid double counting, diagrams for the self-energies must be constrained to the skeleton set in all the dressed channels: they must remain connected after cutting any two lines of the same kind; an example is shown in Fig.~\ref{fig4}.
In addition, one has to enforce two rules concerning dots:
(1) no two dots can be connected directly by two Green's functions, and (2) a dot cannot be connected by two Green's functions to the same end of $G_{pp}$ or $G_{ph}$ or
$W$. This is necessary because, by construction, such diagrams are already accounted
for in the corresponding ladder sums. Finally, there is one exception to the rule: to avoid triple counting of the same diagram
contributing to the lowest-order $\Sigma_{\sigma}$, one has to perform subtraction of the diagram based on two
points, see Fig.~\ref{fig5}.

All our results are based on the lowest-order GGGW scheme, in which the self-energies are given by the skeleton diagrams shown in Fig.~\ref{fig5} and the lines are computed self-consistently by Eqs.~(\ref{Dysons1}), (\ref{Dysons2}). In principle, higher-order skeleton graphs can be summed by the standard diagrammatic Monte Carlo (DiagMC) scheme \cite{van2010diagrammatic, kozik2010diagrammatic, kulagin2,deng2015emergent} with obvious modifications required to handle a larger set of diagrammatic elements and self energies. We have implemented the DiagMC scheme and used it to assess systematic errors of the lowest-order approximation.

Formally, the exact answer follows from summing the whole infinite series of all skeleton diagrams for the self-energies. However, it is known that in strongly correlated regimes close to half-filling ($U \gtrsim 4$, $T \lesssim 0.5$, $n \sim 1$) skeleton sums for the Hubbard model cannot converge to the correct answer being attracted to an unphysical branch of the Luttinger-Ward functional~\cite{kozik2015nonexistence}. On the other hand, at weaker interactions skeleton series quickly converge to the exact solution. Therefore, by continuity it is natural to expect that a low-order skeleton theory produces qualitatively and even quantitatively accurate results somewhat into the strongly correlated regime, provided the self-consistent dressing adequately captures fluctuations in relevant channels. We observe that it is indeed the case for the first-order GGGW approach employed here by benchmarking the results at $n=1$, which is notoriously the most difficult regime for skeleton schemes~\cite{kozik2015nonexistence}, against the numerically exact determinant diagrammatic Monte Carlo method \cite{burovski2006critical, DDMC_Neel}. In particular, we observe that in the range or interaction strengths considered here, our approach produces accurate thermodynamic observables at the level of a few percent. The functional form of $\Sigma_{\sigma }(\omega_n,\vec{q})$ and $\chi (\omega_n,\vec{q})$ displays all the features and the deviation of the overall amplitude from the exact answer is at most $30\%$. For the purposes of the qualitative analysis carried out below this level of accuracy is sufficient. This makes the lowest-order GGGW scheme a practical computationally-inexpensive tool that captures complex correlation effects in the regime of parameters accessible in current cold-atom experiments.

{\it Results.}
The very notion of $T_{M}$ as the crossover temperature between the high-temperature
regime, in which the magnetic correlation length $\xi$ is less than or comparable to the lattice constant $a$, and the regime of strong long-range correlations with $\xi \gg a$,
implies that its definition is not unique. To characterise the onset of magnetic
correlations we examine the momentum dependence of the static magnetic susceptibility
$\chi (\omega=0,\vec{q})$ and monitor development of a narrow peak structure. We define $T_{M}$ at a given value of interaction $U$ and density $n$ as the highest temperature at which the amplitude of the peak in $\chi (\omega=0,\vec{q})$ is an order of magnitude larger than its minimum value over the Brillouin zone,
$\chi_{\text{max}}/\chi_{\text{min}} = 10$. Our results are summarized in Fig.~\ref{fig2}.

As expected, the largest values of $T_{M}$ (at fixed $U$) are observed at half-filling
where the crossover temperature can be as high as $ T \sim 0.25$ (or about $1000$~K
in units representative of the $CuO_2$ superconductors with hopping amplitute $t \sim 0.3\,eV$ \cite{gull2013superconductivity}).
As a function of interaction, $T_{M}$ goes to zero at both large and small values of $U$, and features
a smooth maximum around $U \sim 4$. This appears to be the optimal spot for experimental studies
of magnetism in the Hubbard model (\ref{Hubbard}) where reaching low temperatures remains challenging.
The magnetic crossover scale eventually goes to zero with doping but remains relatively high for
intermediate values of $U$ even at doping levels $\delta \in (0.15, 0.25)$.
We did not see evidence for PS at $T_{M}$, meaning that the PS dome takes place within the
magnetic region, see Fig.~\ref{fig1}.

\begin{figure}
\centering
\includegraphics[width=0.47\linewidth]{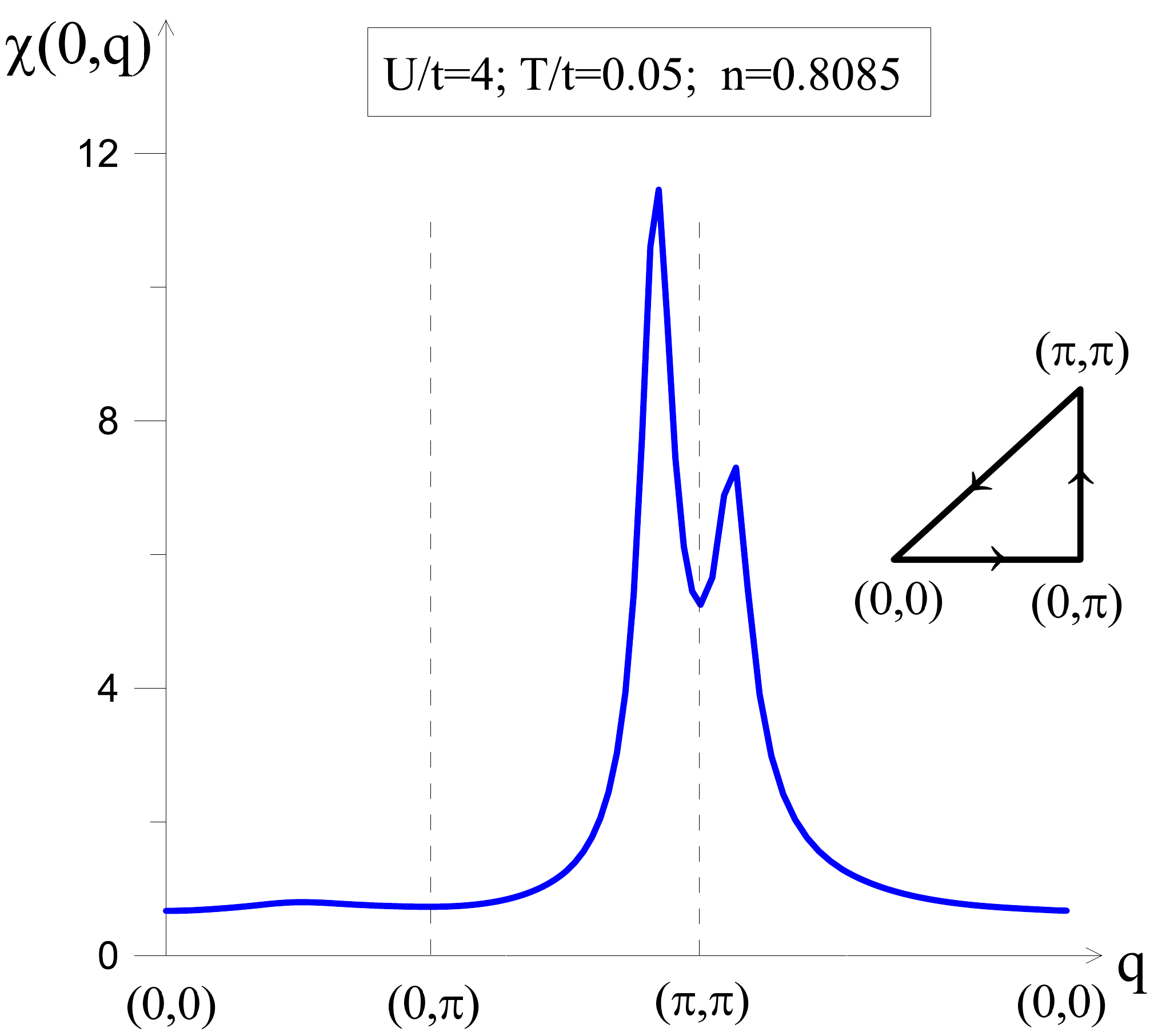}
\hspace{0.0cm}
\includegraphics[width=0.43\linewidth]{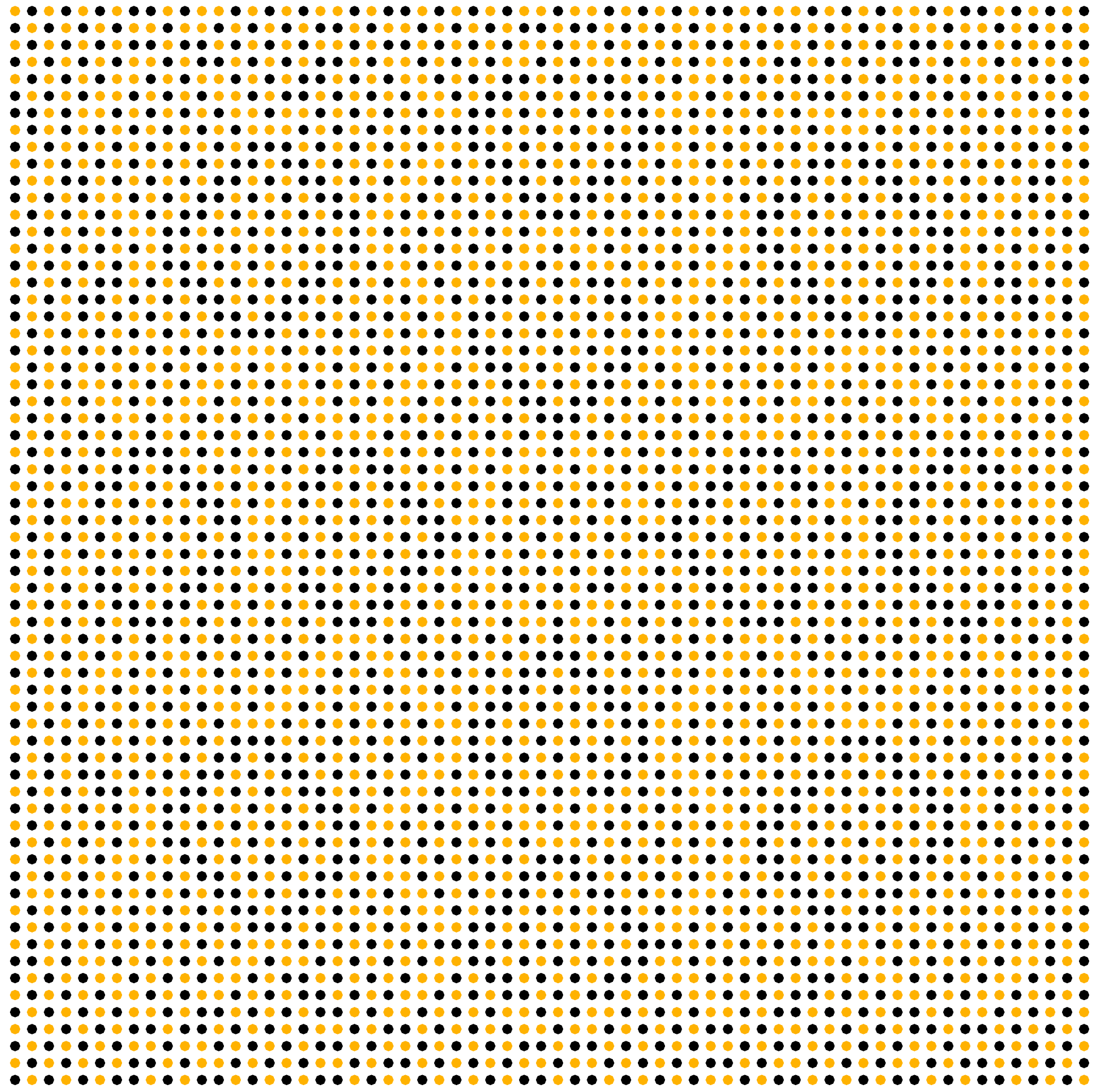}
\caption{\label{fig6} (color online).
\emph{left:} Static magnetic susceptibility from the
lowest-order GGGW scheme plotted over the trace $(0,0) \to (\pi,0) \to (\pi,\pi) \to (0,0)$
in the Brillouin zone for $T=0.05$, $U=4$ and $n=0.8085$.
Two peaks are observed at the incommensurate vectors $\left( \pi, Q\right)$ and $\left( Q, Q\right)$,
where $Q$ is close to $\pi$.
\emph{right:} The corresponding diagonally oriented ``carpet" spin texture is plotted in real space
with colors reflecting the sign of the average spin correlation. Whether this pattern survives
or changes as $T\to 0$ remains an open question.}
\end{figure}

The character of spin correlations undergoes a dramatic transformation with doping.
A mismatch between the largest momentum transfer at the Fermi surface and the
reciprocal lattice vector results in the incommensurate spin-wave fluctuations that take the
form of AFM domains with {\it diagonal} domain walls (or ``diagonal carpet", for brevity).
In Fig.~\ref{fig6}, we show a typical example of the emerging spin structure
(for $T=0.05$, $U=4$ and $n=0.8085$). In the left panel, we see that the otherwise dominant peak
around the commensurate vector $\left( \pi, \pi \right)$ is split and features a minimum at $\left( \pi, \pi \right)$ surrounded by two maxima at the incommensurate vectors. The real-space spin texture
behind this split-peak signal
is shown in the right panel with different colors corresponding
to the sign of $\chi(0,\vec{r})$. It is plausible
that in the PS region, see Fig.~\ref{fig1}, the AFM order is intermixed with the diagonal carpet,
and the mechanism for the $d_{x^2-y^2}$-wave pairing is based on coupling to these spin fluctuations.

{\it Conclusions.}
We discussed the finite-temperature phase diagram of the repulsive Fermi Hubbard model
on a square lattice and identified the overarching dome defining the onset of strong
magnetic correlations that change their structure from commensurate antiferromagnetism
to incommensurate diagonal texture as the doping level is increased. Given relatively
high values of $T_{M}$ that extend well into the doping region where optimal values
for transition temperatures to the  $d_{x^2-y^2}$ superfluidity are expected to take place,
magnetic correlations appear to be the key ingredient behind both the PS and superfluidity near half-filling.

Further development of the Diagrammatic Monte Carlo approach is required to
obtain controllable results at temperatures below $T_{M}$ where convergence of
the diagrammatic expansion becomes problematic. Large magnetic correlation length and the possibility of
phase separation should be treated with extreme care by
any numerical method based on finite-cluster calculations because this physics imposes
restrictions on the minimal acceptable cluster size and questions
homogeneity of the solution. In particular, the superfluid states proposed in \cite{gull2013superconductivity,chen2015superconducting,zheng2016ground} in the region of parameter space where one can expect phase separation \cite{chang2008spatially,sorella16,tocchio16} could result from the intrinsic bias of computational technique. To find high-$T_c$ regions one has to avoid PS near half-filling by doping or
modify the model to include non-zero values of the next-nearest-neighbor hopping $t^{\prime}>0$
\cite{chen2015superconducting}.

Ultra-cold atom experiments are expected to have a major impact on
revealing the finite-temperature phase diagram. Current experiments
have already reached temperatures $T \sim T_{M}$ \cite{greinerprivate},
and are well positioned to explore the structure of strong magnetic correlations.
Detection and characterization of the PS state requires reaching lower temperature scales. However, there is no \textit{a priori} reason for the PS dome to take place at $T \ll T_{M}$ given
that known correlations saturate quickly below $T_{M}$.

Quantifying magnetic correlations is also of significant interest in relation with copper oxides \cite{kastner1998magnetic}
as neutron scattering experiments have revealed the coexistence of commensurate and incommensurate
magnetic structures at finite doping. For $\text{La}_{2-p}\text{Sr}_p\text{Cu}_4$ an incommensurate state with a magnetic structure wave vector was found at small dopings; for $\text{YBa}_2\text{Cu}_3\text{O}_{6+y}$ a wide doping window exists where commensurate AFM fluctuations are observed at low temperatures \cite{matsuda2002electronic}.

The authors are grateful to Shiwei Zhang, Sandro Sorella, Luca F. Tocchio, and Markus Greiner
for useful discussions of their results.
Fruitful exchanges with Ulrich Schollw\"ock about the possibility of phase separation at larger $U$'s are acknowledged. Fedor \v{S}imkovic would like to thank USTC Hefei for generous hospitality during a period when parts of this paper were written.
This work was supported by the Simons Collaboration on the Many Electron Problem, National Science Foundation under the grant PHY-1314735, the MURI Program ``New Quantum Phases of Matter" from AFOSR, and the Swiss National Science Foundation, NSFC Grant No.~11625522, CAS, and NKBRSFC Grant No.~2016YFA0301600.

\bibliography{refs}

\begin{thebibliography}{52}%
\makeatletter
\providecommand \@ifxundefined [1]{%
 \@ifx{#1\undefined}
}%
\providecommand \@ifnum [1]{%
 \ifnum #1\expandafter \@firstoftwo
 \else \expandafter \@secondoftwo
 \fi
}%
\providecommand \@ifx [1]{%
 \ifx #1\expandafter \@firstoftwo
 \else \expandafter \@secondoftwo
 \fi
}%
\providecommand \natexlab [1]{#1}%
\providecommand \enquote  [1]{``#1''}%
\providecommand \bibnamefont  [1]{#1}%
\providecommand \bibfnamefont [1]{#1}%
\providecommand \citenamefont [1]{#1}%
\providecommand \href@noop [0]{\@secondoftwo}%
\providecommand \href [0]{\begingroup \@sanitize@url \@href}%
\providecommand \@href[1]{\@@startlink{#1}\@@href}%
\providecommand \@@href[1]{\endgroup#1\@@endlink}%
\providecommand \@sanitize@url [0]{\catcode `\\12\catcode `\$12\catcode
  `\&12\catcode `\#12\catcode `\^12\catcode `\_12\catcode `\%12\relax}%
\providecommand \@@startlink[1]{}%
\providecommand \@@endlink[0]{}%
\providecommand \url  [0]{\begingroup\@sanitize@url \@url }%
\providecommand \@url [1]{\endgroup\@href {#1}{\urlprefix }}%
\providecommand \urlprefix  [0]{URL }%
\providecommand \Eprint [0]{\href }%
\providecommand \doibase [0]{http://dx.doi.org/}%
\providecommand \selectlanguage [0]{\@gobble}%
\providecommand \bibinfo  [0]{\@secondoftwo}%
\providecommand \bibfield  [0]{\@secondoftwo}%
\providecommand \translation [1]{[#1]}%
\providecommand \BibitemOpen [0]{}%
\providecommand \bibitemStop [0]{}%
\providecommand \bibitemNoStop [0]{.\EOS\space}%
\providecommand \EOS [0]{\spacefactor3000\relax}%
\providecommand \BibitemShut  [1]{\csname bibitem#1\endcsname}%
\let\auto@bib@innerbib\@empty
\bibitem [{\citenamefont {Hubbard}(1963)}]{hubbard1963electron}%
  \BibitemOpen
  \bibfield  {author} {\bibinfo {author} {\bibfnamefont {J.}~\bibnamefont
  {Hubbard}},\ }in\ \href@noop {} {\emph {\bibinfo {booktitle} {Proceedings of
  the royal society of london a: mathematical, physical and engineering
  sciences}}},\ Vol.\ \bibinfo {volume} {276}\ (\bibinfo {organization} {The
  Royal Society},\ \bibinfo {year} {1963})\ pp.\ \bibinfo {pages}
  {238--257}\BibitemShut {NoStop}%
\bibitem [{\citenamefont {Anderson}(1963)}]{anderson1963theory}%
  \BibitemOpen
  \bibfield  {author} {\bibinfo {author} {\bibfnamefont {P.~W.}\ \bibnamefont
  {Anderson}},\ }\href@noop {} {\bibfield  {journal} {\bibinfo  {journal}
  {Solid state physics}\ }\textbf {\bibinfo {volume} {14}},\ \bibinfo {pages}
  {99} (\bibinfo {year} {1963})}\BibitemShut {NoStop}%
\bibitem [{\citenamefont {Anderson}\ \emph {et~al.}(1997)\citenamefont
  {Anderson} \emph {et~al.}}]{anderson1997theory}%
  \BibitemOpen
  \bibfield  {author} {\bibinfo {author} {\bibfnamefont {P.~W.}\ \bibnamefont
  {Anderson}} \emph {et~al.},\ }\href@noop {} {\emph {\bibinfo {title} {The
  theory of superconductivity in the high-Tc cuprate superconductors}}},\ Vol.\
  \bibinfo {volume} {446}\ (\bibinfo  {publisher} {Princeton University Press
  Princeton, NJ},\ \bibinfo {year} {1997})\BibitemShut {NoStop}%
\bibitem [{\citenamefont {Lee}\ \emph {et~al.}(2006)\citenamefont {Lee},
  \citenamefont {Nagaosa},\ and\ \citenamefont {Wen}}]{Lee2006}%
  \BibitemOpen
  \bibfield  {author} {\bibinfo {author} {\bibfnamefont {P.~A.}\ \bibnamefont
  {Lee}}, \bibinfo {author} {\bibfnamefont {N.}~\bibnamefont {Nagaosa}}, \ and\
  \bibinfo {author} {\bibfnamefont {X.-G.}\ \bibnamefont {Wen}},\ }\href
  {\doibase 10.1103/RevModPhys.78.17} {\bibfield  {journal} {\bibinfo
  {journal} {Rev. Mod. Phys.}\ }\textbf {\bibinfo {volume} {78}},\ \bibinfo
  {pages} {17} (\bibinfo {year} {2006})}\BibitemShut {NoStop}%
\bibitem [{\citenamefont {Jaksch}\ \emph {et~al.}(1998)\citenamefont {Jaksch},
  \citenamefont {Bruder}, \citenamefont {Cirac}, \citenamefont {Gardiner},\
  and\ \citenamefont {Zoller}}]{jaksch1998cold}%
  \BibitemOpen
  \bibfield  {author} {\bibinfo {author} {\bibfnamefont {D.}~\bibnamefont
  {Jaksch}}, \bibinfo {author} {\bibfnamefont {C.}~\bibnamefont {Bruder}},
  \bibinfo {author} {\bibfnamefont {J.~I.}\ \bibnamefont {Cirac}}, \bibinfo
  {author} {\bibfnamefont {C.~W.}\ \bibnamefont {Gardiner}}, \ and\ \bibinfo
  {author} {\bibfnamefont {P.}~\bibnamefont {Zoller}},\ }\href@noop {}
  {\bibfield  {journal} {\bibinfo  {journal} {Physical Review Letters}\
  }\textbf {\bibinfo {volume} {81}},\ \bibinfo {pages} {3108} (\bibinfo {year}
  {1998})}\BibitemShut {NoStop}%
\bibitem [{\citenamefont {Lewenstein}\ \emph {et~al.}(2007)\citenamefont
  {Lewenstein}, \citenamefont {Sanpera}, \citenamefont {Ahufinger},
  \citenamefont {Damski}, \citenamefont {Sen},\ and\ \citenamefont
  {Sen}}]{lewenstein2007ultracold}%
  \BibitemOpen
  \bibfield  {author} {\bibinfo {author} {\bibfnamefont {M.}~\bibnamefont
  {Lewenstein}}, \bibinfo {author} {\bibfnamefont {A.}~\bibnamefont {Sanpera}},
  \bibinfo {author} {\bibfnamefont {V.}~\bibnamefont {Ahufinger}}, \bibinfo
  {author} {\bibfnamefont {B.}~\bibnamefont {Damski}}, \bibinfo {author}
  {\bibfnamefont {A.}~\bibnamefont {Sen}}, \ and\ \bibinfo {author}
  {\bibfnamefont {U.}~\bibnamefont {Sen}},\ }\href@noop {} {\bibfield
  {journal} {\bibinfo  {journal} {Advances in Physics}\ }\textbf {\bibinfo
  {volume} {56}},\ \bibinfo {pages} {243} (\bibinfo {year} {2007})}\BibitemShut
  {NoStop}%
\bibitem [{\citenamefont {K{\"o}hl}\ \emph {et~al.}(2005)\citenamefont
  {K{\"o}hl}, \citenamefont {Moritz}, \citenamefont {St{\"o}ferle},
  \citenamefont {G{\"u}nter},\ and\ \citenamefont
  {Esslinger}}]{kohl2005fermionic}%
  \BibitemOpen
  \bibfield  {author} {\bibinfo {author} {\bibfnamefont {M.}~\bibnamefont
  {K{\"o}hl}}, \bibinfo {author} {\bibfnamefont {H.}~\bibnamefont {Moritz}},
  \bibinfo {author} {\bibfnamefont {T.}~\bibnamefont {St{\"o}ferle}}, \bibinfo
  {author} {\bibfnamefont {K.}~\bibnamefont {G{\"u}nter}}, \ and\ \bibinfo
  {author} {\bibfnamefont {T.}~\bibnamefont {Esslinger}},\ }\href@noop {}
  {\bibfield  {journal} {\bibinfo  {journal} {Physical Review Letters}\
  }\textbf {\bibinfo {volume} {94}},\ \bibinfo {pages} {080403} (\bibinfo
  {year} {2005})}\BibitemShut {NoStop}%
\bibitem [{\citenamefont {J{\"o}rdens}\ \emph {et~al.}(2008)\citenamefont
  {J{\"o}rdens}, \citenamefont {Strohmaier}, \citenamefont {G{\"u}nter},
  \citenamefont {Moritz},\ and\ \citenamefont {Esslinger}}]{jordens2008mott}%
  \BibitemOpen
  \bibfield  {author} {\bibinfo {author} {\bibfnamefont {R.}~\bibnamefont
  {J{\"o}rdens}}, \bibinfo {author} {\bibfnamefont {N.}~\bibnamefont
  {Strohmaier}}, \bibinfo {author} {\bibfnamefont {K.}~\bibnamefont
  {G{\"u}nter}}, \bibinfo {author} {\bibfnamefont {H.}~\bibnamefont {Moritz}},
  \ and\ \bibinfo {author} {\bibfnamefont {T.}~\bibnamefont {Esslinger}},\
  }\href@noop {} {\bibfield  {journal} {\bibinfo  {journal} {Nature}\ }\textbf
  {\bibinfo {volume} {455}},\ \bibinfo {pages} {204} (\bibinfo {year}
  {2008})}\BibitemShut {NoStop}%
\bibitem [{\citenamefont {Schneider}\ \emph {et~al.}(2008)\citenamefont
  {Schneider}, \citenamefont {Hackerm{\"u}ller}, \citenamefont {Will},
  \citenamefont {Best}, \citenamefont {Bloch}, \citenamefont {Costi},
  \citenamefont {Helmes}, \citenamefont {Rasch},\ and\ \citenamefont
  {Rosch}}]{schneider2008metallic}%
  \BibitemOpen
  \bibfield  {author} {\bibinfo {author} {\bibfnamefont {U.}~\bibnamefont
  {Schneider}}, \bibinfo {author} {\bibfnamefont {L.}~\bibnamefont
  {Hackerm{\"u}ller}}, \bibinfo {author} {\bibfnamefont {S.}~\bibnamefont
  {Will}}, \bibinfo {author} {\bibfnamefont {T.}~\bibnamefont {Best}}, \bibinfo
  {author} {\bibfnamefont {I.}~\bibnamefont {Bloch}}, \bibinfo {author}
  {\bibfnamefont {T.}~\bibnamefont {Costi}}, \bibinfo {author} {\bibfnamefont
  {R.}~\bibnamefont {Helmes}}, \bibinfo {author} {\bibfnamefont
  {D.}~\bibnamefont {Rasch}}, \ and\ \bibinfo {author} {\bibfnamefont
  {A.}~\bibnamefont {Rosch}},\ }\href@noop {} {\bibfield  {journal} {\bibinfo
  {journal} {Science}\ }\textbf {\bibinfo {volume} {322}},\ \bibinfo {pages}
  {1520} (\bibinfo {year} {2008})}\BibitemShut {NoStop}%
\bibitem [{\citenamefont {Hulet}\ \emph {et~al.}(2015)\citenamefont {Hulet},
  \citenamefont {Duarte}, \citenamefont {Hart},\ and\ \citenamefont
  {Yang}}]{hulet2015antiferromagnetism}%
  \BibitemOpen
  \bibfield  {author} {\bibinfo {author} {\bibfnamefont {R.~G.}\ \bibnamefont
  {Hulet}}, \bibinfo {author} {\bibfnamefont {P.~M.}\ \bibnamefont {Duarte}},
  \bibinfo {author} {\bibfnamefont {R.~A.}\ \bibnamefont {Hart}}, \ and\
  \bibinfo {author} {\bibfnamefont {T.-L.}\ \bibnamefont {Yang}},\ }\href@noop
  {} {\bibfield  {journal} {\bibinfo  {journal} {arXiv preprint
  arXiv:1512.05311}\ } (\bibinfo {year} {2015})}\BibitemShut {NoStop}%
\bibitem [{\citenamefont {Greif}\ \emph {et~al.}(2015)\citenamefont {Greif},
  \citenamefont {Jotzu}, \citenamefont {Messer}, \citenamefont {Desbuquois},\
  and\ \citenamefont {Esslinger}}]{greif2015formation}%
  \BibitemOpen
  \bibfield  {author} {\bibinfo {author} {\bibfnamefont {D.}~\bibnamefont
  {Greif}}, \bibinfo {author} {\bibfnamefont {G.}~\bibnamefont {Jotzu}},
  \bibinfo {author} {\bibfnamefont {M.}~\bibnamefont {Messer}}, \bibinfo
  {author} {\bibfnamefont {R.}~\bibnamefont {Desbuquois}}, \ and\ \bibinfo
  {author} {\bibfnamefont {T.}~\bibnamefont {Esslinger}},\ }\href@noop {}
  {\bibfield  {journal} {\bibinfo  {journal} {Physical Review Letters}\
  }\textbf {\bibinfo {volume} {115}},\ \bibinfo {pages} {260401} (\bibinfo
  {year} {2015})}\BibitemShut {NoStop}%
\bibitem [{\citenamefont {Parsons}\ \emph {et~al.}(2016)\citenamefont
  {Parsons}, \citenamefont {Mazurenko}, \citenamefont {Chiu}, \citenamefont
  {Ji}, \citenamefont {Greif},\ and\ \citenamefont
  {Greiner}}]{parsons2016site}%
  \BibitemOpen
  \bibfield  {author} {\bibinfo {author} {\bibfnamefont {M.~F.}\ \bibnamefont
  {Parsons}}, \bibinfo {author} {\bibfnamefont {A.}~\bibnamefont {Mazurenko}},
  \bibinfo {author} {\bibfnamefont {C.~S.}\ \bibnamefont {Chiu}}, \bibinfo
  {author} {\bibfnamefont {G.}~\bibnamefont {Ji}}, \bibinfo {author}
  {\bibfnamefont {D.}~\bibnamefont {Greif}}, \ and\ \bibinfo {author}
  {\bibfnamefont {M.}~\bibnamefont {Greiner}},\ }\href@noop {} {\bibfield
  {journal} {\bibinfo  {journal} {arXiv preprint arXiv:1605.02704}\ } (\bibinfo
  {year} {2016})}\BibitemShut {NoStop}%
\bibitem [{\citenamefont {Mazurenko}\ \emph {et~al.}()\citenamefont
  {Mazurenko}, \citenamefont {Chiu}, \citenamefont {Ji}, \citenamefont
  {Parsons}, \citenamefont {Kanász-Nagy}, \citenamefont {Schmidt},
  \citenamefont {Grusdt}, \citenamefont {Demler}, \citenamefont {Greif},\ and\
  \citenamefont {Greiner}}]{greinerprivate}%
  \BibitemOpen
  \bibfield  {author} {\bibinfo {author} {\bibfnamefont {A.}~\bibnamefont
  {Mazurenko}}, \bibinfo {author} {\bibfnamefont {C.~S.}\ \bibnamefont {Chiu}},
  \bibinfo {author} {\bibfnamefont {G.}~\bibnamefont {Ji}}, \bibinfo {author}
  {\bibfnamefont {M.~F.}\ \bibnamefont {Parsons}}, \bibinfo {author}
  {\bibfnamefont {M.}~\bibnamefont {Kanász-Nagy}}, \bibinfo {author}
  {\bibfnamefont {R.}~\bibnamefont {Schmidt}}, \bibinfo {author} {\bibfnamefont
  {F.}~\bibnamefont {Grusdt}}, \bibinfo {author} {\bibfnamefont
  {E.}~\bibnamefont {Demler}}, \bibinfo {author} {\bibfnamefont
  {D.}~\bibnamefont {Greif}}, \ and\ \bibinfo {author} {\bibfnamefont
  {M.}~\bibnamefont {Greiner}},\ }\href@noop {} {\bibinfo  {journal}
  {arXiv:1612.08436}\ }\BibitemShut {NoStop}%
\bibitem [{\citenamefont {Kagan}\ and\ \citenamefont
  {Chubukov}(1989)}]{kagan1989increase}%
  \BibitemOpen
\bibfield  {journal} {  }\bibfield  {author} {\bibinfo {author} {\bibfnamefont
  {M.~Y.}\ \bibnamefont {Kagan}}\ and\ \bibinfo {author} {\bibfnamefont
  {A.}~\bibnamefont {Chubukov}},\ }\href@noop {} {\bibfield  {journal}
  {\bibinfo  {journal} {Soviet Journal of Experimental and Theoretical Physics
  Letters}\ }\textbf {\bibinfo {volume} {50}},\ \bibinfo {pages} {517}
  (\bibinfo {year} {1989})}\BibitemShut {NoStop}%
\bibitem [{\citenamefont {Baranov}\ and\ \citenamefont
  {Kagan}(1992)}]{baranov1992d}%
  \BibitemOpen
  \bibfield  {author} {\bibinfo {author} {\bibfnamefont {M.}~\bibnamefont
  {Baranov}}\ and\ \bibinfo {author} {\bibfnamefont {M.~Y.}\ \bibnamefont
  {Kagan}},\ }\href@noop {} {\bibfield  {journal} {\bibinfo  {journal}
  {Zeitschrift f{\"u}r Physik B Condensed Matter}\ }\textbf {\bibinfo {volume}
  {86}},\ \bibinfo {pages} {237} (\bibinfo {year} {1992})}\BibitemShut
  {NoStop}%
\bibitem [{\citenamefont {Chubukov}\ and\ \citenamefont
  {Lu}(1992)}]{chubukov1992pairing}%
  \BibitemOpen
  \bibfield  {author} {\bibinfo {author} {\bibfnamefont {A.~V.}\ \bibnamefont
  {Chubukov}}\ and\ \bibinfo {author} {\bibfnamefont {J.~P.}\ \bibnamefont
  {Lu}},\ }\href@noop {} {\bibfield  {journal} {\bibinfo  {journal} {Physical
  Review B}\ }\textbf {\bibinfo {volume} {46}},\ \bibinfo {pages} {11163}
  (\bibinfo {year} {1992})}\BibitemShut {NoStop}%
\bibitem [{\citenamefont {Chubukov}(1993)}]{chubukov1993kohn}%
  \BibitemOpen
  \bibfield  {author} {\bibinfo {author} {\bibfnamefont {A.~V.}\ \bibnamefont
  {Chubukov}},\ }\href@noop {} {\bibfield  {journal} {\bibinfo  {journal}
  {Physical Review B}\ }\textbf {\bibinfo {volume} {48}},\ \bibinfo {pages}
  {1097} (\bibinfo {year} {1993})}\BibitemShut {NoStop}%
\bibitem [{\citenamefont {Zanchi}\ and\ \citenamefont
  {Schulz}(1996)}]{zanchi1996superconducting}%
  \BibitemOpen
  \bibfield  {author} {\bibinfo {author} {\bibfnamefont {D.}~\bibnamefont
  {Zanchi}}\ and\ \bibinfo {author} {\bibfnamefont {H.}~\bibnamefont
  {Schulz}},\ }\href@noop {} {\bibfield  {journal} {\bibinfo  {journal}
  {Physical Review B}\ }\textbf {\bibinfo {volume} {54}},\ \bibinfo {pages}
  {9509} (\bibinfo {year} {1996})}\BibitemShut {NoStop}%
\bibitem [{\citenamefont {Halboth}\ and\ \citenamefont
  {Metzner}(2000)}]{halboth2000d}%
  \BibitemOpen
  \bibfield  {author} {\bibinfo {author} {\bibfnamefont {C.~J.}\ \bibnamefont
  {Halboth}}\ and\ \bibinfo {author} {\bibfnamefont {W.}~\bibnamefont
  {Metzner}},\ }\href@noop {} {\bibfield  {journal} {\bibinfo  {journal}
  {Physical Review Letters}\ }\textbf {\bibinfo {volume} {85}},\ \bibinfo
  {pages} {5162} (\bibinfo {year} {2000})}\BibitemShut {NoStop}%
\bibitem [{\citenamefont {Fukazawa}\ and\ \citenamefont
  {Yamada}(2002)}]{fukazawa2002third}%
  \BibitemOpen
  \bibfield  {author} {\bibinfo {author} {\bibfnamefont {H.}~\bibnamefont
  {Fukazawa}}\ and\ \bibinfo {author} {\bibfnamefont {K.}~\bibnamefont
  {Yamada}},\ }\href@noop {} {\bibfield  {journal} {\bibinfo  {journal}
  {Journal of the Physical Society of Japan}\ }\textbf {\bibinfo {volume}
  {71}},\ \bibinfo {pages} {1541} (\bibinfo {year} {2002})}\BibitemShut
  {NoStop}%
\bibitem [{\citenamefont {Hlubina}(1999)}]{hlubina1999phase}%
  \BibitemOpen
  \bibfield  {author} {\bibinfo {author} {\bibfnamefont {R.}~\bibnamefont
  {Hlubina}},\ }\href@noop {} {\bibfield  {journal} {\bibinfo  {journal}
  {Physical Review B}\ }\textbf {\bibinfo {volume} {59}},\ \bibinfo {pages}
  {9600} (\bibinfo {year} {1999})}\BibitemShut {NoStop}%
\bibitem [{\citenamefont {Raghu}\ \emph {et~al.}(2010)\citenamefont {Raghu},
  \citenamefont {Kivelson},\ and\ \citenamefont
  {Scalapino}}]{raghu2010superconductivity}%
  \BibitemOpen
  \bibfield  {author} {\bibinfo {author} {\bibfnamefont {S.}~\bibnamefont
  {Raghu}}, \bibinfo {author} {\bibfnamefont {S.}~\bibnamefont {Kivelson}}, \
  and\ \bibinfo {author} {\bibfnamefont {D.}~\bibnamefont {Scalapino}},\
  }\href@noop {} {\bibfield  {journal} {\bibinfo  {journal} {Physical Review
  B}\ }\textbf {\bibinfo {volume} {81}},\ \bibinfo {pages} {224505} (\bibinfo
  {year} {2010})}\BibitemShut {NoStop}%
\bibitem [{\citenamefont {R{\o}mer}\ \emph {et~al.}(2015)\citenamefont
  {R{\o}mer}, \citenamefont {Kreisel}, \citenamefont {Eremin}, \citenamefont
  {Malakhov}, \citenamefont {Maier}, \citenamefont {Hirschfeld},\ and\
  \citenamefont {Andersen}}]{romer2015pairing}%
  \BibitemOpen
  \bibfield  {author} {\bibinfo {author} {\bibfnamefont {A.~T.}\ \bibnamefont
  {R{\o}mer}}, \bibinfo {author} {\bibfnamefont {A.}~\bibnamefont {Kreisel}},
  \bibinfo {author} {\bibfnamefont {I.}~\bibnamefont {Eremin}}, \bibinfo
  {author} {\bibfnamefont {M.}~\bibnamefont {Malakhov}}, \bibinfo {author}
  {\bibfnamefont {T.}~\bibnamefont {Maier}}, \bibinfo {author} {\bibfnamefont
  {P.}~\bibnamefont {Hirschfeld}}, \ and\ \bibinfo {author} {\bibfnamefont
  {B.}~\bibnamefont {Andersen}},\ }\href@noop {} {\bibfield  {journal}
  {\bibinfo  {journal} {Physical Review B}\ }\textbf {\bibinfo {volume} {92}},\
  \bibinfo {pages} {104505} (\bibinfo {year} {2015})}\BibitemShut {NoStop}%
\bibitem [{\citenamefont {{\v{S}}imkovic~IV}\ \emph {et~al.}(2016)\citenamefont
  {{\v{S}}imkovic~IV}, \citenamefont {Liu}, \citenamefont {Deng},\ and\
  \citenamefont {Kozik}}]{simkovic2016ground}%
  \BibitemOpen
  \bibfield  {author} {\bibinfo {author} {\bibfnamefont {F.}~\bibnamefont
  {{\v{S}}imkovic~IV}}, \bibinfo {author} {\bibfnamefont {X.-W.}\ \bibnamefont
  {Liu}}, \bibinfo {author} {\bibfnamefont {Y.}~\bibnamefont {Deng}}, \ and\
  \bibinfo {author} {\bibfnamefont {E.}~\bibnamefont {Kozik}},\ }\href@noop {}
  {\bibfield  {journal} {\bibinfo  {journal} {Physical Review B}\ }\textbf
  {\bibinfo {volume} {94}},\ \bibinfo {pages} {085106} (\bibinfo {year}
  {2016})}\BibitemShut {NoStop}%
\bibitem [{\citenamefont {Deng}\ \emph {et~al.}(2015)\citenamefont {Deng},
  \citenamefont {Kozik}, \citenamefont {Prokof'ev},\ and\ \citenamefont
  {Svistunov}}]{deng2015emergent}%
  \BibitemOpen
  \bibfield  {author} {\bibinfo {author} {\bibfnamefont {Y.}~\bibnamefont
  {Deng}}, \bibinfo {author} {\bibfnamefont {E.}~\bibnamefont {Kozik}},
  \bibinfo {author} {\bibfnamefont {N.~V.}\ \bibnamefont {Prokof'ev}}, \ and\
  \bibinfo {author} {\bibfnamefont {B.~V.}\ \bibnamefont {Svistunov}},\
  }\href@noop {} {\bibfield  {journal} {\bibinfo  {journal} {EPL (Europhysics
  Letters)}\ }\textbf {\bibinfo {volume} {110}},\ \bibinfo {pages} {57001}
  (\bibinfo {year} {2015})}\BibitemShut {NoStop}%
\bibitem [{\citenamefont {Hirsch}(1985)}]{hirsch1985two}%
  \BibitemOpen
  \bibfield  {author} {\bibinfo {author} {\bibfnamefont {J.~E.}\ \bibnamefont
  {Hirsch}},\ }\href@noop {} {\bibfield  {journal} {\bibinfo  {journal}
  {Physical Review B}\ }\textbf {\bibinfo {volume} {31}},\ \bibinfo {pages}
  {4403} (\bibinfo {year} {1985})}\BibitemShut {NoStop}%
\bibitem [{\citenamefont {Hirsch}\ and\ \citenamefont
  {Tang}(1989)}]{hirsch1989antiferromagnetism}%
  \BibitemOpen
  \bibfield  {author} {\bibinfo {author} {\bibfnamefont {J.}~\bibnamefont
  {Hirsch}}\ and\ \bibinfo {author} {\bibfnamefont {S.}~\bibnamefont {Tang}},\
  }\href@noop {} {\bibfield  {journal} {\bibinfo  {journal} {Physical Review
  Letters}\ }\textbf {\bibinfo {volume} {62}},\ \bibinfo {pages} {591}
  (\bibinfo {year} {1989})}\BibitemShut {NoStop}%
\bibitem [{\citenamefont {White}\ \emph {et~al.}(1989)\citenamefont {White},
  \citenamefont {Scalapino}, \citenamefont {Sugar}, \citenamefont {Loh},
  \citenamefont {Gubernatis},\ and\ \citenamefont
  {Scalettar}}]{white1989numerical}%
  \BibitemOpen
  \bibfield  {author} {\bibinfo {author} {\bibfnamefont {S.}~\bibnamefont
  {White}}, \bibinfo {author} {\bibfnamefont {D.}~\bibnamefont {Scalapino}},
  \bibinfo {author} {\bibfnamefont {R.}~\bibnamefont {Sugar}}, \bibinfo
  {author} {\bibfnamefont {E.}~\bibnamefont {Loh}}, \bibinfo {author}
  {\bibfnamefont {J.}~\bibnamefont {Gubernatis}}, \ and\ \bibinfo {author}
  {\bibfnamefont {R.}~\bibnamefont {Scalettar}},\ }\href@noop {} {\bibfield
  {journal} {\bibinfo  {journal} {Physical Review B}\ }\textbf {\bibinfo
  {volume} {40}},\ \bibinfo {pages} {506} (\bibinfo {year} {1989})}\BibitemShut
  {NoStop}%
\bibitem [{\citenamefont {Furukawa}\ and\ \citenamefont
  {Imada}(1992)}]{furukawa1992two}%
  \BibitemOpen
  \bibfield  {author} {\bibinfo {author} {\bibfnamefont {N.}~\bibnamefont
  {Furukawa}}\ and\ \bibinfo {author} {\bibfnamefont {M.}~\bibnamefont
  {Imada}},\ }\href@noop {} {\bibfield  {journal} {\bibinfo  {journal} {Journal
  of the Physical Society of Japan}\ }\textbf {\bibinfo {volume} {61}},\
  \bibinfo {pages} {3331} (\bibinfo {year} {1992})}\BibitemShut {NoStop}%
\bibitem [{\citenamefont {Varney}\ \emph {et~al.}(2009)\citenamefont {Varney},
  \citenamefont {Lee}, \citenamefont {Bai}, \citenamefont {Chiesa},
  \citenamefont {Jarrell},\ and\ \citenamefont
  {Scalettar}}]{varney2009quantum}%
  \BibitemOpen
  \bibfield  {author} {\bibinfo {author} {\bibfnamefont {C.}~\bibnamefont
  {Varney}}, \bibinfo {author} {\bibfnamefont {C.-R.}\ \bibnamefont {Lee}},
  \bibinfo {author} {\bibfnamefont {Z.}~\bibnamefont {Bai}}, \bibinfo {author}
  {\bibfnamefont {S.}~\bibnamefont {Chiesa}}, \bibinfo {author} {\bibfnamefont
  {M.}~\bibnamefont {Jarrell}}, \ and\ \bibinfo {author} {\bibfnamefont
  {R.}~\bibnamefont {Scalettar}},\ }\href@noop {} {\bibfield  {journal}
  {\bibinfo  {journal} {Physical Review B}\ }\textbf {\bibinfo {volume} {80}},\
  \bibinfo {pages} {075116} (\bibinfo {year} {2009})}\BibitemShut {NoStop}%
\bibitem [{\citenamefont {Penn}(1966)}]{penn1966stability}%
  \BibitemOpen
  \bibfield  {author} {\bibinfo {author} {\bibfnamefont {D.~R.}\ \bibnamefont
  {Penn}},\ }\href@noop {} {\bibfield  {journal} {\bibinfo  {journal} {Physical
  Review}\ }\textbf {\bibinfo {volume} {142}},\ \bibinfo {pages} {350}
  (\bibinfo {year} {1966})}\BibitemShut {NoStop}%
\bibitem [{\citenamefont {Nagaoka}(1966)}]{nagaoka1966ferromagnetism}%
  \BibitemOpen
  \bibfield  {author} {\bibinfo {author} {\bibfnamefont {Y.}~\bibnamefont
  {Nagaoka}},\ }\href@noop {} {\bibfield  {journal} {\bibinfo  {journal}
  {Physical Review}\ }\textbf {\bibinfo {volume} {147}},\ \bibinfo {pages}
  {392} (\bibinfo {year} {1966})}\BibitemShut {NoStop}%
\bibitem [{\citenamefont {Tasaki}(1998)}]{tasaki1998nagaoka}%
  \BibitemOpen
  \bibfield  {author} {\bibinfo {author} {\bibfnamefont {H.}~\bibnamefont
  {Tasaki}},\ }\href@noop {} {\bibfield  {journal} {\bibinfo  {journal}
  {Progress of Theoretical Physics}\ }\textbf {\bibinfo {volume} {99}},\
  \bibinfo {pages} {489} (\bibinfo {year} {1998})}\BibitemShut {NoStop}%
\bibitem [{\citenamefont {Zitzler}\ \emph {et~al.}(2002)\citenamefont
  {Zitzler}, \citenamefont {Pruschke},\ and\ \citenamefont
  {Bulla}}]{zitzler2002magnetism}%
  \BibitemOpen
  \bibfield  {author} {\bibinfo {author} {\bibfnamefont {R.}~\bibnamefont
  {Zitzler}}, \bibinfo {author} {\bibfnamefont {T.}~\bibnamefont {Pruschke}}, \
  and\ \bibinfo {author} {\bibfnamefont {R.}~\bibnamefont {Bulla}},\
  }\href@noop {} {\bibfield  {journal} {\bibinfo  {journal} {The European
  Physical Journal B-Condensed Matter and Complex Systems}\ }\textbf {\bibinfo
  {volume} {27}},\ \bibinfo {pages} {473} (\bibinfo {year} {2002})}\BibitemShut
  {NoStop}%
\bibitem [{\citenamefont {Schulz}(1990)}]{schulz1990incommensurate}%
  \BibitemOpen
  \bibfield  {author} {\bibinfo {author} {\bibfnamefont {H.}~\bibnamefont
  {Schulz}},\ }\href@noop {} {\bibfield  {journal} {\bibinfo  {journal}
  {Physical Review Letters}\ }\textbf {\bibinfo {volume} {64}},\ \bibinfo
  {pages} {1445} (\bibinfo {year} {1990})}\BibitemShut {NoStop}%
\bibitem [{\citenamefont {Lin}(1991)}]{lin1991ground}%
  \BibitemOpen
  \bibfield  {author} {\bibinfo {author} {\bibfnamefont {H.}~\bibnamefont
  {Lin}},\ }\href@noop {} {\bibfield  {journal} {\bibinfo  {journal} {Physical
  Review B}\ }\textbf {\bibinfo {volume} {44}},\ \bibinfo {pages} {7151}
  (\bibinfo {year} {1991})}\BibitemShut {NoStop}%
\bibitem [{\citenamefont {Chang}\ and\ \citenamefont
  {Zhang}(2008)}]{chang2008spatially}%
  \BibitemOpen
  \bibfield  {author} {\bibinfo {author} {\bibfnamefont {C.-C.}\ \bibnamefont
  {Chang}}\ and\ \bibinfo {author} {\bibfnamefont {S.}~\bibnamefont {Zhang}},\
  }\href@noop {} {\bibfield  {journal} {\bibinfo  {journal} {Physical Review
  B}\ }\textbf {\bibinfo {volume} {78}},\ \bibinfo {pages} {165101} (\bibinfo
  {year} {2008})}\BibitemShut {NoStop}%
\bibitem [{\citenamefont {Sorella}(2015)}]{sorella16}%
  \BibitemOpen
  \bibfield  {author} {\bibinfo {author} {\bibfnamefont {S.}~\bibnamefont
  {Sorella}},\ }\href {\doibase 10.1103/PhysRevB.91.241116} {\bibfield
  {journal} {\bibinfo  {journal} {Phys. Rev. B}\ }\textbf {\bibinfo {volume}
  {91}},\ \bibinfo {pages} {241116} (\bibinfo {year} {2015})}\BibitemShut
  {NoStop}%
\bibitem [{\citenamefont {Tocchio}\ \emph {et~al.}(2016)\citenamefont
  {Tocchio}, \citenamefont {Becca},\ and\ \citenamefont {Sorella}}]{tocchio16}%
  \BibitemOpen
  \bibfield  {author} {\bibinfo {author} {\bibfnamefont {L.~F.}\ \bibnamefont
  {Tocchio}}, \bibinfo {author} {\bibfnamefont {F.}~\bibnamefont {Becca}}, \
  and\ \bibinfo {author} {\bibfnamefont {S.}~\bibnamefont {Sorella}},\ }\href
  {\doibase 10.1103/PhysRevB.94.195126} {\bibfield  {journal} {\bibinfo
  {journal} {Phys. Rev. B}\ }\textbf {\bibinfo {volume} {94}},\ \bibinfo
  {pages} {195126} (\bibinfo {year} {2016})}\BibitemShut {NoStop}%
\bibitem [{\citenamefont {Abrikosov}\ \emph {et~al.}(1975)\citenamefont
  {Abrikosov}, \citenamefont {Gor'kov},\ and\ \citenamefont
  {Dzyaloshinski}}]{AGD}%
  \BibitemOpen
  \bibfield  {author} {\bibinfo {author} {\bibfnamefont {A.~A.}\ \bibnamefont
  {Abrikosov}}, \bibinfo {author} {\bibfnamefont {L.~P.}\ \bibnamefont
  {Gor'kov}}, \ and\ \bibinfo {author} {\bibfnamefont {I.~E.}\ \bibnamefont
  {Dzyaloshinski}},\ }\href@noop {} {\emph {\bibinfo {title} {Methods of
  Quantum Field Theory in Statistical Physics}}}\ (\bibinfo  {publisher} {Dover
  Publications Inc.},\ \bibinfo {year} {1975})\BibitemShut {NoStop}%
\bibitem [{Note1()}]{Note1}%
  \BibitemOpen
  \bibinfo {note} {Bare interaction needs to be treated separately to define
  precise rules for avoiding double counting.}\BibitemShut {Stop}%
\bibitem [{\citenamefont {Van~Houcke}\ \emph {et~al.}(2010)\citenamefont
  {Van~Houcke}, \citenamefont {Kozik}, \citenamefont {Prokof’ev},\ and\
  \citenamefont {Svistunov}}]{van2010diagrammatic}%
  \BibitemOpen
  \bibfield  {author} {\bibinfo {author} {\bibfnamefont {K.}~\bibnamefont
  {Van~Houcke}}, \bibinfo {author} {\bibfnamefont {E.}~\bibnamefont {Kozik}},
  \bibinfo {author} {\bibfnamefont {N.}~\bibnamefont {Prokof’ev}}, \ and\
  \bibinfo {author} {\bibfnamefont {B.}~\bibnamefont {Svistunov}},\ }\href@noop
  {} {\bibfield  {journal} {\bibinfo  {journal} {Physics Procedia}\ }\textbf
  {\bibinfo {volume} {6}},\ \bibinfo {pages} {95} (\bibinfo {year}
  {2010})}\BibitemShut {NoStop}%
\bibitem [{\citenamefont {Kozik}\ \emph {et~al.}(2010)\citenamefont {Kozik},
  \citenamefont {Van~Houcke}, \citenamefont {Gull}, \citenamefont {Pollet},
  \citenamefont {Prokof'ev}, \citenamefont {Svistunov},\ and\ \citenamefont
  {Troyer}}]{kozik2010diagrammatic}%
  \BibitemOpen
  \bibfield  {author} {\bibinfo {author} {\bibfnamefont {E.}~\bibnamefont
  {Kozik}}, \bibinfo {author} {\bibfnamefont {K.}~\bibnamefont {Van~Houcke}},
  \bibinfo {author} {\bibfnamefont {E.}~\bibnamefont {Gull}}, \bibinfo {author}
  {\bibfnamefont {L.}~\bibnamefont {Pollet}}, \bibinfo {author} {\bibfnamefont
  {N.}~\bibnamefont {Prokof'ev}}, \bibinfo {author} {\bibfnamefont
  {B.}~\bibnamefont {Svistunov}}, \ and\ \bibinfo {author} {\bibfnamefont
  {M.}~\bibnamefont {Troyer}},\ }\href@noop {} {\bibfield  {journal} {\bibinfo
  {journal} {EPL (Europhysics Letters)}\ }\textbf {\bibinfo {volume} {90}},\
  \bibinfo {pages} {10004} (\bibinfo {year} {2010})}\BibitemShut {NoStop}%
\bibitem [{\citenamefont {Kulagin}\ \emph {et~al.}(2013)\citenamefont
  {Kulagin}, \citenamefont {Prokof'ev}, \citenamefont {Starykh}, \citenamefont
  {Svistunov},\ and\ \citenamefont {Varney}}]{kulagin2}%
  \BibitemOpen
  \bibfield  {author} {\bibinfo {author} {\bibfnamefont {S.~A.}\ \bibnamefont
  {Kulagin}}, \bibinfo {author} {\bibfnamefont {N.}~\bibnamefont {Prokof'ev}},
  \bibinfo {author} {\bibfnamefont {O.~A.}\ \bibnamefont {Starykh}}, \bibinfo
  {author} {\bibfnamefont {B.}~\bibnamefont {Svistunov}}, \ and\ \bibinfo
  {author} {\bibfnamefont {C.~N.}\ \bibnamefont {Varney}},\ }\href {\doibase
  10.1103/PhysRevB.87.024407} {\bibfield  {journal} {\bibinfo  {journal} {Phys.
  Rev. B}\ }\textbf {\bibinfo {volume} {87}},\ \bibinfo {pages} {024407}
  (\bibinfo {year} {2013})}\BibitemShut {NoStop}%
\bibitem [{\citenamefont {Kozik}\ \emph {et~al.}(2015)\citenamefont {Kozik},
  \citenamefont {Ferrero},\ and\ \citenamefont
  {Georges}}]{kozik2015nonexistence}%
  \BibitemOpen
  \bibfield  {author} {\bibinfo {author} {\bibfnamefont {E.}~\bibnamefont
  {Kozik}}, \bibinfo {author} {\bibfnamefont {M.}~\bibnamefont {Ferrero}}, \
  and\ \bibinfo {author} {\bibfnamefont {A.}~\bibnamefont {Georges}},\
  }\href@noop {} {\bibfield  {journal} {\bibinfo  {journal} {Physical Review
  Letters}\ }\textbf {\bibinfo {volume} {114}},\ \bibinfo {pages} {156402}
  (\bibinfo {year} {2015})}\BibitemShut {NoStop}%
\bibitem [{\citenamefont {Burovski}\ \emph {et~al.}(2006)\citenamefont
  {Burovski}, \citenamefont {Prokof’ev}, \citenamefont {Svistunov},\ and\
  \citenamefont {Troyer}}]{burovski2006critical}%
  \BibitemOpen
  \bibfield  {author} {\bibinfo {author} {\bibfnamefont {E.}~\bibnamefont
  {Burovski}}, \bibinfo {author} {\bibfnamefont {N.}~\bibnamefont
  {Prokof’ev}}, \bibinfo {author} {\bibfnamefont {B.}~\bibnamefont
  {Svistunov}}, \ and\ \bibinfo {author} {\bibfnamefont {M.}~\bibnamefont
  {Troyer}},\ }\href@noop {} {\bibfield  {journal} {\bibinfo  {journal}
  {Physical Review Letters}\ }\textbf {\bibinfo {volume} {96}},\ \bibinfo
  {pages} {160402} (\bibinfo {year} {2006})}\BibitemShut {NoStop}%
\bibitem [{\citenamefont {Kozik}\ \emph {et~al.}(2013)\citenamefont {Kozik},
  \citenamefont {Burovski}, \citenamefont {Scarola},\ and\ \citenamefont
  {Troyer}}]{DDMC_Neel}%
  \BibitemOpen
  \bibfield  {author} {\bibinfo {author} {\bibfnamefont {E.}~\bibnamefont
  {Kozik}}, \bibinfo {author} {\bibfnamefont {E.}~\bibnamefont {Burovski}},
  \bibinfo {author} {\bibfnamefont {V.~W.}\ \bibnamefont {Scarola}}, \ and\
  \bibinfo {author} {\bibfnamefont {M.}~\bibnamefont {Troyer}},\ }\href
  {\doibase 10.1103/PhysRevB.87.205102} {\bibfield  {journal} {\bibinfo
  {journal} {Phys. Rev. B}\ }\textbf {\bibinfo {volume} {87}},\ \bibinfo
  {pages} {205102} (\bibinfo {year} {2013})}\BibitemShut {NoStop}%
\bibitem [{\citenamefont {Gull}\ \emph {et~al.}(2013)\citenamefont {Gull},
  \citenamefont {Parcollet},\ and\ \citenamefont
  {Millis}}]{gull2013superconductivity}%
  \BibitemOpen
  \bibfield  {author} {\bibinfo {author} {\bibfnamefont {E.}~\bibnamefont
  {Gull}}, \bibinfo {author} {\bibfnamefont {O.}~\bibnamefont {Parcollet}}, \
  and\ \bibinfo {author} {\bibfnamefont {A.~J.}\ \bibnamefont {Millis}},\
  }\href@noop {} {\bibfield  {journal} {\bibinfo  {journal} {Physical Review
  Letters}\ }\textbf {\bibinfo {volume} {110}},\ \bibinfo {pages} {216405}
  (\bibinfo {year} {2013})}\BibitemShut {NoStop}%
\bibitem [{\citenamefont {Chen}\ \emph {et~al.}(2015)\citenamefont {Chen},
  \citenamefont {LeBlanc},\ and\ \citenamefont
  {Gull}}]{chen2015superconducting}%
  \BibitemOpen
  \bibfield  {author} {\bibinfo {author} {\bibfnamefont {X.}~\bibnamefont
  {Chen}}, \bibinfo {author} {\bibfnamefont {J.}~\bibnamefont {LeBlanc}}, \
  and\ \bibinfo {author} {\bibfnamefont {E.}~\bibnamefont {Gull}},\ }\href@noop
  {} {\bibfield  {journal} {\bibinfo  {journal} {Physical Review Letters}\
  }\textbf {\bibinfo {volume} {115}},\ \bibinfo {pages} {116402} (\bibinfo
  {year} {2015})}\BibitemShut {NoStop}%
\bibitem [{\citenamefont {Zheng}\ and\ \citenamefont
  {Chan}(2016)}]{zheng2016ground}%
  \BibitemOpen
  \bibfield  {author} {\bibinfo {author} {\bibfnamefont {B.-X.}\ \bibnamefont
  {Zheng}}\ and\ \bibinfo {author} {\bibfnamefont {G.~K.-L.}\ \bibnamefont
  {Chan}},\ }\href@noop {} {\bibfield  {journal} {\bibinfo  {journal} {Physical
  Review B}\ }\textbf {\bibinfo {volume} {93}},\ \bibinfo {pages} {035126}
  (\bibinfo {year} {2016})}\BibitemShut {NoStop}%
\bibitem [{\citenamefont {Kastner}\ \emph {et~al.}(1998)\citenamefont
  {Kastner}, \citenamefont {Birgeneau}, \citenamefont {Shirane},\ and\
  \citenamefont {Endoh}}]{kastner1998magnetic}%
  \BibitemOpen
  \bibfield  {author} {\bibinfo {author} {\bibfnamefont {M.}~\bibnamefont
  {Kastner}}, \bibinfo {author} {\bibfnamefont {R.}~\bibnamefont {Birgeneau}},
  \bibinfo {author} {\bibfnamefont {G.}~\bibnamefont {Shirane}}, \ and\
  \bibinfo {author} {\bibfnamefont {Y.}~\bibnamefont {Endoh}},\ }\href@noop {}
  {\bibfield  {journal} {\bibinfo  {journal} {Reviews of Modern Physics}\
  }\textbf {\bibinfo {volume} {70}},\ \bibinfo {pages} {897} (\bibinfo {year}
  {1998})}\BibitemShut {NoStop}%
\bibitem [{\citenamefont {Matsuda}\ \emph {et~al.}(2002)\citenamefont
  {Matsuda}, \citenamefont {Fujita}, \citenamefont {Yamada}, \citenamefont
  {Birgeneau}, \citenamefont {Endoh},\ and\ \citenamefont
  {Shirane}}]{matsuda2002electronic}%
  \BibitemOpen
  \bibfield  {author} {\bibinfo {author} {\bibfnamefont {M.}~\bibnamefont
  {Matsuda}}, \bibinfo {author} {\bibfnamefont {M.}~\bibnamefont {Fujita}},
  \bibinfo {author} {\bibfnamefont {K.}~\bibnamefont {Yamada}}, \bibinfo
  {author} {\bibfnamefont {R.}~\bibnamefont {Birgeneau}}, \bibinfo {author}
  {\bibfnamefont {Y.}~\bibnamefont {Endoh}}, \ and\ \bibinfo {author}
  {\bibfnamefont {G.}~\bibnamefont {Shirane}},\ }\href@noop {} {\bibfield
  {journal} {\bibinfo  {journal} {Physical Review B}\ }\textbf {\bibinfo
  {volume} {65}},\ \bibinfo {pages} {134515} (\bibinfo {year}
  {2002})}\BibitemShut {NoStop}%
\end{thebibliography}%

\end{document}